\documentclass[twocolumn,showpacs,preprintnumbers,amsmath,amssymb]{revtex4}

\pdfoutput=1
\usepackage{xspace}
\usepackage{epsfig}
\usepackage{amsfonts}
\usepackage{graphicx}
\usepackage{url}
\usepackage[ 
              colorlinks,%
          linkcolor=blue,%
        citecolor = blue,%
              hyperindex,%
        plainpages=false,%
         urlcolor = blue,%
       pdfstartview=FitH]{hyperref}

\begin{document}


\title{\textsf{ Mobility in Semiconducting Graphene Nanoribbons: Phonon, Impurity, \& Edge Roughness Scattering }}

\author{Tian Fang}
\author{Aniruddha Konar}
\author{Huili Xing}
\author{Debdeep Jena}
\email[Electronic mail: ]{djena@nd.edu}

\affiliation{Department of Electrical Engineering, University of Notre Dame, IN, 46556, USA}

\date{\today}

\begin{abstract}
The transport properties of carriers in semiconducting graphene nanoribbons are studied by comparing the effects of phonon, impurity, and line-edge roughness scattering.  It is found that scattering from impurities located at the surface of nanoribbons, and from acoustic phonons are as important as line edge roughness scattering.  The relative importance of these scattering mechanisms varies with the temperature, Fermi level location, and the width of the ribbons.  Based on the analysis, strategies for improvement of low-field mobility are described.
\end{abstract}



\maketitle
\section{Introduction} 

Since the isolation of 2-dimensional (2D) graphene sheets in 2004 \cite{geim07natmat, novo05nat, science06gatech}, there has been substantial interest in patterning them into quasi-1D graphene nanoribbons (GNRs) \cite{prl07kim}.  This interest is motivated by the ability to tune the bandgap of GNRs by lithographic processes, and the consequent freedom it is expected to afford in lithographic bandgap-engineered electronic devices.  GNRs as thin as $\sim3$ nm have been achieved, exhibiting bandgaps approaching 0.3 - 0.4 eV, which makes them attractive for many low-power device applications \cite{science08_gnr_fets}.  The measured carrier mobilities in the first reported ultrathin GNRs are reported to be much lower \cite{prl08_dai_gnr_mobility} than in corresponding 2D graphene sheets \cite{prl08_gr_mobility}.  For improvements in the transport properties, it is essential to identify the major sources of carrier scattering in thin GNRs.  To that end, this work presents a comprehensive analysis of various scattering mechanisms in GNRs.  Analytical scattering rates for phonon, impurity, and line-edge roughness (LER) scattering are derived taking the GNR wavefunctions into account.  It is shown that carrier mobilities in GNRs are not necessarily limited by edge roughness scattering; similar to the case in 2D graphene, surface impurity \cite{prB07_gr_transport} and phonon scattering play important roles, depending on the carrier concentrations, widths, and temperature.  The results of this work are expected to provide useful strategies towards improvements in carrier transport properties in GNRs for various applications.

The paper is structured in the following fashion.  Section II describes the theoretical formalism used for the scattering rate and mobility calculations, with a discussion of the electronic structure, scattering matrix elements, and screening in GNRs.  In section III, individual scattering rates due to acoustic and optical phonons, line edge roughness (LER), and bulk and surface impurities are derived.  Finally, in section IV, the resulting mobility in GNRs is calculated, with a discussion of the relative importance of the various scattering mechanisms.  The effect of temperature, location of Fermi level, and GNR width on the transport properties is evaluated in that section.  

\section{Theoretical Formalism}\label{formalism}  

Under the application of a small electric field along the GNR axis, the carrier distribution function in the relaxation time approximation is given by
\begin{equation}
f \approx f_{0} - (e \tau v_{g} \mathcal{F} ) \cdot ( - \frac{ \partial f_{0} }{ \partial \mathcal E } ),
\end{equation}
where $\tau^{-1}$ is the scattering rate in the diffusive limit, $v_{g}$ is the group velocity of carriers, ${\mathcal F}$ is the electric field, $e$ is the electron charge, and $f_{0}$ is the equilibrium Fermi-Dirac distribution.  If the scattering rate is known, the current flowing through the GNR may be evaluated by the relation 
\begin{equation}
I = e (g_{s} g_{v}/L) \sum_{k} f v_{g}, 
\end{equation} 
where $e$ is the electron charge, $L$ is the GNR length, $g_{s}=2$ is the spin degeneracy and $g_{v} = 1$ is the valley degeneracy of semiconducting GNRs.  Noting that the electric field is $\mathcal{F} = V/L$ where $V$ is the applied voltage, the 1D conductivity (in units of S-m) is given by
\begin{equation}
\sigma_{1D} =  \frac{2 e^{2}}{h} \int_{0}^{\infty}  v_{g}(\mathcal{E}) \tau(\mathcal{E}) ( - \frac{ \partial f_{0} }{ \partial \mathcal E } ) d\mathcal{E},
\end{equation}
where $h = 2 \pi \hbar$ is Planck's constant.  The mobility is calculated as $\mu = \sigma_{1D}/ e n_{1D} $, where $n_{1D}$ is the one-dimensional carrier density.  This formalism allows for the evaluation of scattering rates, mobility, and conductivity for a general Fermi level location, which can be assumed to be tuned capacitively through a gate voltage in experiments, at any temperature.

\subsection{Electronic Structure} Semiconducting armchair GNRs with lengths $L (>> W)$ along the $y$ ($x$) directions are considered in this work.  The hard-wall boundary conditions at the edges lead to a bandstructure $\mathcal{E} ( k_{n} , k_{y}) = \hbar v_{F} \sqrt{ k_{n}^{2} + k_{y}^{2} }$, where $v_{F} \sim $10$^{8}$ cm/s is the Fermi velocity.  The allowed transverse wavevector for semiconducting armchair GNRs are quantized to values $k_{n} = \pm {n \pi}/{3W}$, depending on the width $W$; here $n = \pm 1, \pm 2, \pm 4, \pm 5, \pm 7, \pm 8, ...$.  The resulting bandgap is $\mathcal{E}_{g} = 2 \pi \hbar v_{F} / 3  W $, which is $\sim 1.38/W$ eV with $W$ expressed in nm.  We note that the group velocity of carriers in a GNR is linked to the density of states (DOS) $\rho_{GNR}$ by $v_{g} (\mathcal{E})  = 2 / \pi \hbar \rho_{GNR}(\mathcal{E})$, where the DOS is $\rho_{GNR} (\mathcal E) =  (2/ \pi \hbar v_{F}) \times ( \mathcal{E}/ \sqrt{ \mathcal{E}^{2} - (\mathcal{E}_{g}/2)^{2} } ) $ for the first subband.  The effective 1D carrier density in the GNR is given by $n_{1D} = \int_{0}^{\infty} \rho_{GNR}(\mathcal{E}) f_{0} (\mathcal{E}) d\mathcal{E}$.


\subsection{Matrix Elements and Scattering Rates}  

The electronic wavefunctions of carriers in semiconducting GNRs may be derived from an admixture of states in the $\mathcal{K}$ and $\mathcal{K^{\prime}}$ valleys of the underlying 2D graphene bandstructure \cite{PrB06_gnr_bands}.  For armchair semiconducting GNRs, the valley degeneracy is removed.  Assuming a length $L$ and a width $W$, the electronic wavefunctions in the nth subband in the Ket-notation are written as
\begin{equation}
|\varphi\rangle=\frac{1}{\sqrt{2}}|k_{n},k_{y}\rangle-\frac{1}{\sqrt{2}}|\widetilde{k_{n}},k_{y}\rangle,
\end{equation}
where the tilde sign indicates the state in the $\mathcal{K}^{\prime}$ valley. The resulting energy dispersion (the $\mathcal{E} - k$ relation) of the nth subband is $\mathcal{E} (k_{n} , k_{y}) = \hbar  v_{F} \sqrt{ k_{n}^2 + k_{y}^2 } $, where $v_{F} \sim 10^8 $ cm/s is the Fermi velocity in graphene.  The projections of these states on to the real space gives the wavefunctions
\begin{eqnarray}
\lefteqn{
\langle {\bf r} | k_{n}, k_{y} \rangle = \sqrt{ \frac{1}{2LW}} e^{ik_{y}y} e^{i(\frac{\Delta K}{2}-k_{n})x}
\left(\begin{array}{c} 1\\-e^{i\theta_{n}} \end{array}\right) }
\nonumber\\ &&
\langle {\bf r} |\widetilde{k_{n}},k_{y}\rangle=\sqrt{\frac{1}{2LW}}e^{ik_{y}y}e^{i(k_{n}-\frac{\Delta K}{2})x}
\left(\begin{array}{c} 1\\-e^{i\theta_{n}} \end{array}\right) 
\end{eqnarray}
where ${\bf r} = (x,y)$ is the vector in the x-y plane. $\Delta K=4\pi/3a$ is the distance between two Dirac points in the $k$-space for the underlying 2D graphene bandstructure, $a$ being the lattice constant of graphene. $\frac{\Delta K}{2}-k_{n}$ is the electron wave vector in the confined x-direction for the nth subband states. $\theta_{n}$ is the angle between $k_{y}$ and $k_{n}$, $\theta_{n} = \tan^{-1}(k_{y}/k_{n})$.  Since the carrier states in the $k$-space are close to the Dirac point, $k_{n} << \Delta K$.

For a perturbation of the form $V(x, y, z)$, the pertubation matrix element is given by
\begin{eqnarray}
\lefteqn{\langle\varphi '|V(x,y,z)|\varphi\rangle=\frac{1}{2}\Big\{\langle k_{y}',k_{n'}|V|k_{n},k_{y}\rangle+}
\nonumber \\ 
&& \langle k_{y}',\widetilde{k_{n'}}|V|\widetilde{k_{n}},k_{y}\rangle\Big\}-
\nonumber \\
&& \frac{1}{2}\Big\{\langle k_{y}',\widetilde{k_{n'}}|V|k_{n},k_{y}\rangle+\langle k_{y}',k_{n'}|V|\widetilde{k_{n}},k_{y}\rangle\Big\}.
\end{eqnarray}

For finding the mobility and conductance, the scattering rates due to various scattering mechanisms are evaluated using Fermi's golden rule, in the form 
\begin{equation}
S({\bf k}, {\bf k}^{\prime}) = \frac{2 \pi}{\hbar} |  \langle\varphi '|V(x,y,z)|\varphi\rangle |^{2} \delta( \mathcal{E}_{\bf k} - \mathcal{E}_{{\bf k}}^{\prime} \pm \hbar \omega ),
\end{equation}
where the delta function ensures energy conservation for both elastic ($\omega = 0$) inelastic ($\omega \neq 0 $) scattering processes.  The ensemble scattering rate that contributes to the conductivity and mobility is then evaluated for each scattering mechanism as
\begin{equation}
\frac{1}{\tau} = \sum_{k}S({\bf k}, {\bf k}^{\prime}) ( 1 - \cos \alpha ),
\end{equation}
$\alpha$ is the angle between ${\bf k}$ and ${\bf k}^{\prime}$, and the summation runs over all available final states.


\subsection{Screening}
Screening of the scattering potential by free-carriers in the GNRs modify the scattering rates.  To take this many-body effect into account, the following procedure has been used (see \cite{prb07_breyfertig_scr_gnr}). The static screening is calculated using the random phase approximation (RPA). In the $k-$space, the relation between screened potential and unscreened potential is determined by $V^{uns}(q_{n})=\sum_{m}\epsilon(q_{m},q_{n})V^{scr}(q_{m})$. The screening matrix is given by
\begin{equation}
\epsilon(q_{m},q_{n},q_{y})=\delta_{q_{m}q_{n}}+\frac{e^2}{2\pi\epsilon_{0}\kappa}\mathcal{F}(q_{m},q_{n},q_{y})\mathcal{L}(q_{n},q_{y}),
\end{equation}
where $\epsilon_{0}$ is the permittivity of vacuum, $\kappa$ is the average of the dielectric constants of the regions between which the GNR is sandwiched, and $\delta_{q_{m}q_{n}}$ is the Kronecker sign. $\mathcal{F}()$ is a measure of the contribution to screening from coupled transverse modes, given by
\begin{eqnarray}
\mathcal{F}(q_{m},q_{n},q_{y})=\frac{1}{2W^2}\int_{0}^{W}\int_{0}^{W}K_{0}(|q_{y}(x-x')|) \times \nonumber \\
\cos{(q_{m}x)}\cos{(q_{n}x')}dxdx',
\end{eqnarray}
where $K_{0}(...)$ is the zero-order modified Bessel function.
$\mathcal{L}(q_{n},q_{y})$ is given by
\begin{equation}
\mathcal{L}(q_{n},q_{y})=\sum_{|\delta k_{nn'}|=q_{n}}\mathcal{L}_{nn'}(q_{y}),
\end{equation}
which is a sum over Lindhard functions $\mathcal{L}_{nn'}$ between two subbands that have a wave vector difference $q_{n}$ in the transverse ($x-$direction).  This sum includes both intersubband and interband (conduction and valence band) contributions.  The Lindhard function is given by
\begin{eqnarray}
\mathcal{L}_{nn'}( q_{y} ) = \frac{g_{s}}{L} \sum_{k_{y}} (1 + ss'\cos \theta_{k k^{\prime}}) \times \nonumber \\
\frac{ f_{n}(k_{y}) - f_{n'}(k_{y} + q_{y}) }{\mathcal{E}_{\bf k}^{\prime} - \mathcal{E}_{{\bf k}} },
\end{eqnarray}
where $s=1$ is for conduction subbands and $s^{\prime} =-1$ is for valence subbands. $g_{s} = 2$ is the spin degeneracy, and $f_{n}( ... )$ is the Fermi-Dirac distribution function in the nth subband.
If $|q_{m}-q_{n}|=l\pi/W$ and $l$ is an odd number, $\mathcal{F} (q_{m},q_{n},q_{y})=0$. This makes the screening calculation very simple for the lowest subbands. The screening of the first subband is $\epsilon_{scr}=\epsilon(0,0,q_{y})$, which is given by
\begin{equation}
\epsilon_{scr} = 1 + \frac{e^{2}}{ 2 \pi \epsilon_{0} \kappa }  \mathcal{F}(0,0,q_{y}) \mathcal{L} (0,q_{y}),
\label{screeningfactor}
\end{equation}
which is equal to 1 if screening is neglected.  In the long-wavelength limit ($q_{y} \rightarrow 0$), this evaluates to $\epsilon_{scr} \approx 1 + (e^{2} / 2 \pi \epsilon_{0} \kappa ) \times \rho_{GNR}( \mathcal{E}_{F} ) \ln (2 / q_{y} W )$, whereas in the metallic limit ($q_{y} W >> 1$), it is given by $\epsilon_{scr} \approx 1 + e^{2}/  \pi^{2} \epsilon_{0} \kappa \hbar v_{F}$.  The general form given in Equation \ref{screeningfactor} is used for the calculations that follow.


\section{Scattering Mechanisms} Low-field carrier transport in GNRs is affected by various scattering mechanisms.  Among them, scattering by acoustic and optical phonons, charged impurities, and edge-roughness scattering are expected to be most effective.  Other mechanisms such as remote optical phonons due to polar coupling with underlying substrates or dielectrics may be present, but are not considered in this work.  Among the scattering mechanisms considered, phonon scattering sets the intrinsic limit on carrier mobilities.  For GNRs with imperfect edges, line-edge roughness (LER) scattering can be rather strong for very thin ribbons.  Unintentional charged impurities present either attached to the GNRs or spatially separated from them (for example embedded in the dielectric surrounding) also degrade the mobility.  These scattering mechanisms are considered in this work; their effects are compared, and their relative importance is qualitatively and quantitatively evaluated.


\subsection{Acoustic Phonons} For acoustic phonon scattering, the zone-center acoustic phonon interaction with carriers result in intra-valley scattering, which is quasi-elastic.  Here we only consider the longitudinal mode since this mode induces higher deformation potential than the out of plane and flexural modes \cite{apl06gnrIntel}.  The perturbation potential introduced by acoustic phonons is given by
\begin{equation}
V_{ac} ( y ) = \sqrt{ \frac{ n^{\pm} \hbar }{ 2 \rho L W \omega_{ac} }} \cdot D_{ac} q_{y} e^{i {q_{y}  y}},
\end{equation}
where $n^{-} = 1/( \exp{( \hbar \omega_{ac} / k_{B}T )} - 1 )$, $n^{+} = 1 + n^{-}$, $\omega_{ac} = v_{s} q_{y}$ is the acoustic phonon frequency, $D_{ac} \sim 16$ eV is the deformation potential of acoustic phonons \cite{apl07_goldsman}, $\rho \sim 7.6 \times 10^{-8}$ g/cm$^{2}$ is the 2D mass density of graphene, and $v_{s} \sim 2 \times 10^{6}$ cm/s is the sound velocity in 2D graphene.  Using the formalism outlined in section \ref{formalism} leads to a scattering matrix element
\begin{equation}
V_{ac} ( q_{y} ) = D_{ac} |q_{y}| (1+e^{i\theta_{kk'}}) \sqrt{\frac{n^{\pm}\hbar}{ 8 \rho LW\omega}} \delta_{q_{y},\delta k_{y}},\end{equation}
where the Kronecker delta function $\delta_{q_{y}, \delta k_{y}}$ ensures the momentum conservation for the carrier + acoustic phonon system.  The resulting intra-subband scattering rate is given by
\begin{eqnarray}
\frac{1}{\tau_{ac}( \mathcal{E} )} = \frac{n_{ph} D_{ac}^2 \mathcal{E} }{\hbar^2 v_{F}^2 \rho v_{s} W}(1+\cos{\theta_{kk'}}),
\end{eqnarray}
where $n_{ph} = n^{+} + n^{-}$ is used to take both absorption and emission of acoustic phonons into account.  Here $\mathcal{E}$ is the energy of carriers measured with respect to the Dirac point. The scattering rate may be re-written in the form
\begin{eqnarray}
\frac{1}{\tau_{ac}( \mathcal{E} )} = \frac{ n_{ph} \pi D_{ac}^2 q_{y}^{2} }{ 4 \rho W \omega_{ac} } \rho_{GNR}( \mathcal{E}) (1+\cos{\theta_{kk'}}),
\end{eqnarray}
where backscattering restricts the value of $|q_{y}|  = 2 |k_{y}|$.  This form highlights the proportionality of the scattering rate to the DOS of the GNR.  The scattering rate is also inversely proportional to the width of the GNR, and becomes more severe for narrower ribbons, similar to the diameter dependence in carbon nanotubes \cite{apl06gnrIntel}.




\subsection{Optical Phonons} For optical phonon scattering, we consider the zone-boundary phonon of energy $\hbar \omega_{LO} \sim 160 $ meV, with an optical deformation potential $D_{op} \sim 1.4 \times 10^{9}$ eV/cm, assumed to be the same as for 2D graphene.  Optical phonon scattering is inelastic; at low bias voltages and small electric fields, optical phonon emission by carriers is energetically forbidden, and absorption is damped by the high energy and low population of LO phonons.  Since the emission of optical phonons is possible only if the kinetic energy of carriers exceeds $\hbar \omega_{LO}$, this mechanism is important only for highly energetic carriers.  

The perturbation potential introduced by LO phonons is given by
\begin{equation}
V_{op} (y) = \sqrt{ \frac{ n^{\pm} \hbar }{ 2 \rho L W \omega_{LO} }} \cdot D_{op}  e^{i  q_{y} y},
\end{equation}
where $n^{-} = 1/( \exp{( \hbar \omega_{LO} / k_{B}T )} - 1 )$, $n^{+} = 1 + n^{-}$, similar to those of acoustic phonons.  The matrix element for optical phonon scattering is given by
\begin{equation}
V_{op} (q_{y}) = D_{op} (1+e^{i\theta_{kk'}}) \sqrt{\frac{n^{\pm}\hbar}{8\rho LW\omega_{LO}}} \delta_{q_{y}, \delta k_{y} },
\end{equation}
and the final scattering rate is given by
\begin{eqnarray}
\frac{1}{\tau_{op}(\mathcal{E})}=\frac{n^{\pm} \pi D_{op}^2} {4 \rho W \omega_{LO}}\rho_{GNR}(\mathcal{E'})(1+\cos{\theta_{kk'}}),
\end{eqnarray}
where the energy of the carrier after the scattering event changes to $\mathcal{E}^{\prime}=\mathcal{E}\pm \hbar \omega_{LO}$. 




\subsection{Line Edge Roughness Scattering} 
\begin{figure}
\begin{center}
\leavevmode \epsfxsize=3.3in \epsffile{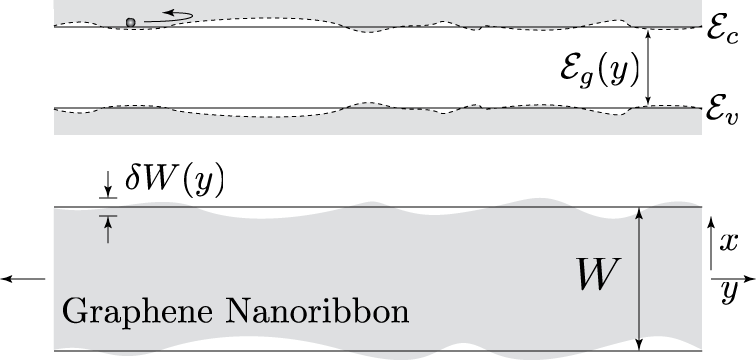}
\caption{Schematic representation of line-edge roughness (LER) scattering in GNRs.  The variation of the GNR width
leads to spatial variation of the bandgap, and the fluctuations in the band edges cause carrier scattering and reduction of
mobility.} 
\label{ler_illustration}
\end{center}
\end{figure}

To capture the effect of line-edge roughness of the GNR on charge transport, the width of ribbon is treated as a function of the longitudinal axis $y$.  The width is given by $ W(y) =  W + \delta W(y)$, where $\delta W(y)$ describes the roughness and $W$ is the spatially averaged width ($< W(y)> = W$ , and  $ < \delta W(y) >  =0$).  This is shown schematically in Figure \ref{ler_illustration}.  The edge roughness $\delta W(y)$ can effectively be described by two parameters and by an exponential spatial correlation function 
\begin{equation}
\langle \delta W(y) \delta W(y + \Delta y) \rangle= H^2 e^{- |\Delta y| / \Lambda },
\end{equation}
where $H$ is the amplitude, and $\Lambda$ is the correlation length of the roughness.  The LER leads to a spatially modulated bandgap, and the resulting fluctuations in the band-edge potential cause the scattering of carriers.  The perturbation potential for the nth subband is given by
\begin{equation}
V_{LER}(y) = -\frac{\delta W(y)}{W} \mathcal{E}_{n},
\label{lerpot}
\end{equation}
where $\mathcal{E}_{n}=\hbar v_{F} |k_{n}|$ is the conduction band energy of the nth subband relative to the Dirac point.  

Since the scattering potential is only dependent on the longitudinal axis, the scattering is intra-subband. The square of matrix element of the nth subband is given by
\begin{equation}
| V_{LER} (q_{y}) | ^2 = \frac{\mathcal{E}_{n}^2}{L} \frac{H^2}{W^2} \frac{\Lambda}{1+(\delta k_{y}\Lambda)^2}(1+\cos\theta_{kk'}),
\end{equation}
leading to a scattering rate 
\begin{equation}
\frac{1}{\tau_{LER}( \mathcal{E} )}=\frac{\pi \mathcal{E}_{n}^2 }{\hbar} \frac{H^2}{W^2} \frac{\Lambda}
{1+4k_{y}^2\Lambda^2}\rho_{GNR}(\mathcal{E})(1+\cos{\theta_{kk'}}).
\end{equation}

Since $\mathcal{E}_{n} \propto 1 / W$, the scattering rate is proportional to $1/W^{4}$, and if the LER scattering rate is dominant, the mobility should scale with the width of the GNR as $\mu_{LER} \sim W^{4}$.  This behavior contrasts with interface/surface roughness scattering in traditional semiconductors with parabolic bandstructures.  Since energy eigenvalues in such traditional semiconductors are proportional to $W^{-2}$, confinement of carriers into a length scale of $W$ leads to a roughness scattering limited mobility which scales as $\mu_{IR} \sim W^{6}$ \cite{apl87_sakaki}.  Thus, the bandstructure of GNRs make them {\em inherently more robust} to LER scattering than parabolic bandgap semiconductor nanostructures of comparable size.

In addition to the the $W^{-4}$ dependence of the LER scattering rate, the dependence on the roughness amplitude is $H^{2}$.  The factor $\Lambda/(1+4k_{y}^2\Lambda^2)$ is maximized for $ 2 k_{y} \sim 1/\Lambda $, indicating that LER scattering is the most severe for those carriers in the GNR that have Fermi wavelengths of the same order as the correlation length of the fluctuations.

\begin{figure}
\begin{center}
\leavevmode \epsfxsize=3.3in \epsffile{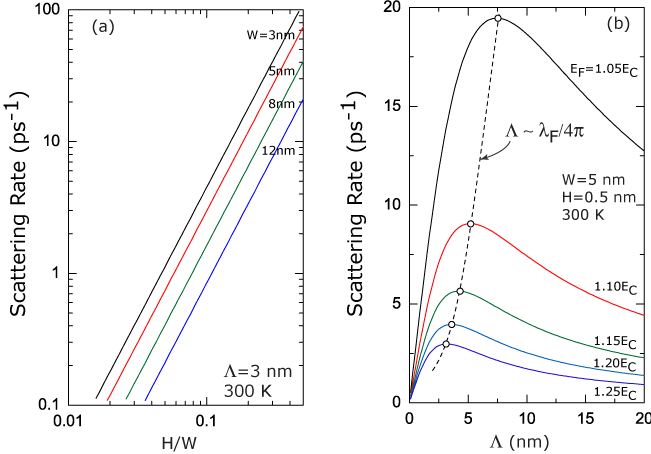}
\caption{Line-edge roughness scattering rates in GNRs for (a) various roughness amplitudes, and (b) various roughness correlation lengths.} 
\label{ler_scatrate}
\end{center}
\end{figure}

The LER scattering rates evaluated for realistic GNRs are plotted in Figure \ref{ler_scatrate} (a) and (b).  Figure \ref{ler_scatrate} (a) shows the LER scattering rate as a function of the roughness amplitude, ranging from $\sim$1\% to $\sim$50\% of the GNR width, for nominal GNR widths $W =$ 3, 5, 8, \& 12 nm.  The correlation length for this plot is $\Lambda \sim 3$ nm.  Figure \ref{ler_scatrate} (b) illustrates the dependence of the scattering rate on the Fermi-level; as the Fermi level in the GNR is increased by a gate voltage, the Fermi-wavelength of carriers decrease, and therefore, LER scattering becomes more sensitive to roughness at shorter correlation lengths.  Therefore, for $2 k_{y} \Lambda << 1$ and for $2 k_{y} \Lambda >> 1$, the LER scattering rate is low.  

We note here that for very rough GNR edges characterized by $H \sim W \sim \Lambda$, it is possible that carriers are localized in quantum-dot like confining potentials; transport in such rough GNRs will occur by hopping and tunneling between localized states as opposed to diffusive band-transport (see \cite{prl07_castroneto}).  The transport treated in this work is restricted to the diffusive band-transport regime.



\subsection{Impurity Scattering} 

\begin{figure}
\begin{center}
\leavevmode \epsfxsize=3.3in \epsffile{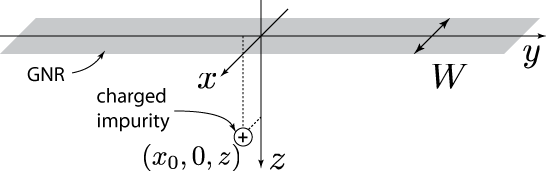}
\caption{Schematic representation of the GNR and charged impurities that cause scattering of mobile carriers.} 
\label{impscatfig}
\end{center}
\end{figure}

Due to an impurity charge located at $ ( x_{0}, 0, z )$ with respect to the GNR which is on the $z$=0 plane (see Figure \ref{impscatfig}), the unscreened scattering potential experienced by a mobile electron located at $( x, y, 0 )$ in the GNR is given by the Coulomb potential 
\begin{equation} 
V_{Coul}(x_{0},y,z) = \frac{e^2}{ 4 \pi \epsilon_{0} \kappa \sqrt{ d^2+y^2 } },
\end{equation}
where $z$ is the distance of the impurity from the GNR plane and $\kappa$ is average relative dielectric constant of materials on the two sides of the GNR \cite{nofringing}.  The distance of the impurity from the origin is $d=\sqrt{ z^2 + ( x - x_{0})^2 }$.  The {\em screened} impurity perturbation matrix element is then given by
\begin{equation}
V_{Coul} (q_{y})  = (1+e^{i\theta_{kk'}}) \frac{e^2}{ 4 \pi \epsilon_{0} \kappa \epsilon_{scr}} \frac{1}{LW} \int_{0}^{W} K_{0} ( |\delta{k_{y}}| d ) dx ,
\end{equation}
where $ \delta{k_{y}} = k_{y} - k_{y'} $ is the change in the carrier wavevector along the GNR axis upon scattering, and $K_{0}()$ is the zeroth order modified Bessel function, and $\epsilon_{scr}$ is the screening factor in Equation \ref{screeningfactor}.

In order to get the scattering rate due to all impurities, we integrate over the distribution of impurities.  For the most general distribution of impurities, the scattering rate is given by
\begin{eqnarray}
\frac{1}{ \tau_{imp} (\mathcal{E})} = \frac{ 2 \pi }{ \hbar }   (\frac{e^2}{4 \pi \epsilon_{0} \kappa \epsilon_{scr} W})^2  \rho_{GNR}(\mathcal{E}) \mathcal{S}(k_{y},W) \times \nonumber \\ 
(1 + \cos{\theta_{kk'}}),
\end{eqnarray}
where $\mathcal{S}$ is an effective scattering `cross-section' with dimensions of length, given by
\begin{equation}
\mathcal{S}(k_{y},W) =  \int_{0}^{t} n_{3d} (z) dz \int_{-\infty}^{\infty} dx_{0} \Big| \int_{0}^{W} K_{0}(2k_{y} d) dx \Big|^2.
\end{equation}
Here $n_{3d}(z)$ is the volume-density of the impurities which can vary with the distance from the GNR plane, $t$ is the thickness over which the impurities are distributed.  This general formalism allows us to evaluate, at the same time, the effect of scattering by volume-distributed impurities as well as impurities located at the GNR surface.  If the charged impurities are located at the GNR/dielectric interface or on the GNR surface, then the scattering rate is found unambiguously by taking $z \rightarrow 0$ and $ n_{3d}(z) \rightarrow n_{2d} \delta( z ) $, where $n_{2d}$ is the 2D impurity density.  For the evaluation of the scattering rates in the rest of the paper, we consider the volume (bulk) and surface impurities separately to highlight their relative importance.  This is motivated by the recent finding that for 2D graphene, surface impurities are responsible for low-field mobility, and when they are removed, an order of magnitude improvement in mobility is observed, even at room temperature \cite{condmat08_kim_susp, condmat08_dassarma_susp}.

\section{Carrier Mobility}

\begin{figure}
\begin{center}
\leavevmode \epsfxsize=2.3in \epsffile{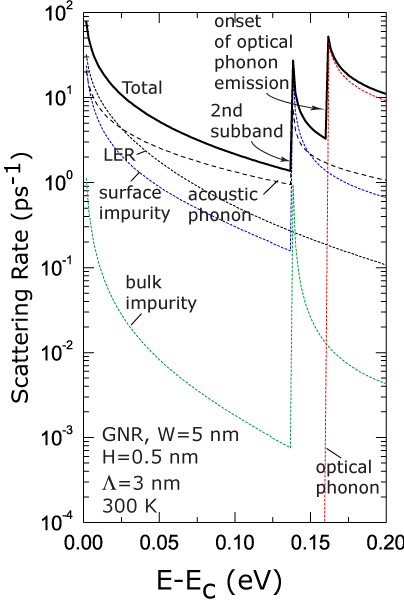}
\caption{Scattering rates due to various scattering mechanisms for a 5 nm wide GNR as a function of the carrier energy.} 
\label{scatrates}
\end{center}
\end{figure}

The rates for each type of scattering mechanism are first evaluated as a function of carrier energy to gauge their relative importance.  They are plotted in Figure \ref{scatrates} for a GNR of $W=5$ nm width at $T= 300$ K.  The relevant parameters used for the calculation are a edge roughness characterized by $(H, \Lambda) = (0.5, 3)$ nm, a bulk impurity density $n_{3d} = 10^{15}$/cm$^{3}$, and a surface impurity density $n_{2d} = 10^{10}$/cm$^{2}$.  In addition to the individual scattering rates, the total scattering rate obtained by Matheissens' rule as a sum of the individual scattering rates is also shown.  For the specific case of the parameters chosen, the LER scattering is seen to dominate at low carrier energies, followed by coustic phonon and surface impurity scattering.  The effect of bulk impurities is found to be relatively weak compared to LER and surface impurities.  It is possible to experimentally lower (or eliminate) scattering due to LER, bulk, and surface impurities, as they are not {\em intrinsic} to the GNR, and such techniques are expected to be developed in due course.  

The intrinsic scattering mechanisms - due to acoustic and optical phonons, are also shown in the figure.  Due to the dependence of the scattering rates on the 1D DOS of the GNR, the general trend is a decrease of scattering rates as the energy of the state increases from the band edge ($\mathcal{E}_{c}$), and then the appearance of a step at the onset of the next subband.  When the energy of a state is $\mathcal{E}_{c} + \hbar \omega_{LO}$, optical phonon emission is allowed, and for such high energy states, optical phonon scattering dominates over all other scattering mechanisms.  This is especially important at high-bias conditions, when high-energy states are occupied.  Optical phonon scattering is responsible for the saturation of current flow through the GNR, leading to a sharp degradation of carrier mobility.  The same has been found earlier for carbon nanotubes \cite{prl05_avouris}.  We restrict the rest of the discussion here to low-bias mobilities.

\begin{figure}
\begin{center}
\leavevmode \epsfxsize=3.3in \epsffile{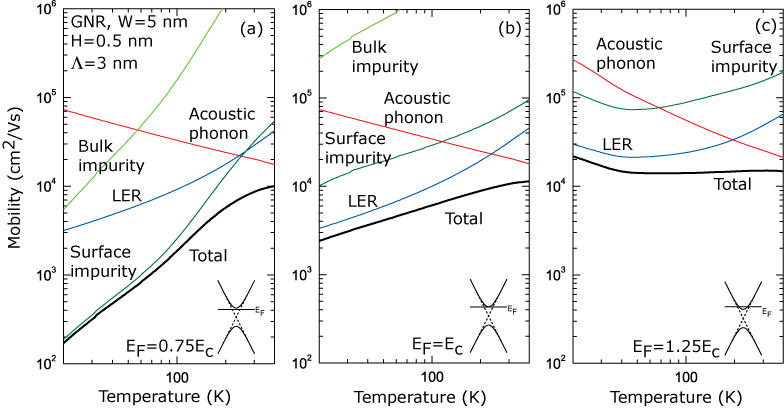}
\caption{Carrier mobility in a 5 nm GNR as a function of temperature for three distinct Fermi levels.  The relative importance of the various scattering rates can be ascertained from the plots.} 
\label{mob5nm}
\end{center}
\end{figure}

Using the formalism for the calculation of carrier mobilities outlined in section \ref{formalism}, the carrier mobility is calculated for a 5 nm wide GNR as a function of temperature for three different locations of the Fermi level.  The impurity densities used in this calculation are $n_{3d} = 10^{15}$/cm$^{3}$ and $n_{2d} = 10^{10}$/cm$^{2}$.  The results are shown in Figure \ref{mob5nm}.  When the Fermi level is located in the bandgap at $\mathcal{E}_{F} = 0.75 \mathcal{E}_{c}$ from the Dirac point, the carrier mobility is severely affected by surface impurity scattering at low temperatures as seen in Figure \ref{mob5nm}(a).  Above $\sim$200 K, acoustic phonon scattering dominates the scattering rate.  The dominance of surface impurity scattering for this case is due to the fact that when $\mathcal{E}_{F} = 0.75 \mathcal{E}_{c}$ and the temperature is low, the net mobile electron density in the conduction band is low, and screening is weak.  As the Fermi level is raised to the conduction band edge - $\mathcal{E}_{F} = \mathcal{E}_{c}$ (Figure \ref{mob5nm}(b)) and then above it - $\mathcal{E}_{F} = 1.25 \mathcal{E}_{c}$ (Figure \ref{mob5nm}(c)), surface Coulomb impurities are strongly screened, and LER scattering dominates at low temperatures.  At and around room temperature, acoustic phonon scattering still limits the electron mobility.  The effect of LER scattering is felt only when the surface impurity density is low.  The surface impurity scattering rate scales inversely with the density of impurities; thus if the surface impurity density is much higher, say $n_{2d} \sim 10^{12}$/cm$^{2}$, it becomes the dominant scattering mechanism, limiting the room temperature mobility to $10 - 1000 $ cm$^{2}$/V.s.  Mobilities in this range has been recently reported in ultrathin GNRs \cite{prl08_dai_gnr_mobility}, but due to the lack of experimental evidence of the GNR edge roughness of such samples, it is early to make a direct comparison.  However, from the analysis, it can be concluded that such values of mobilities are not {\em intrinsic}, and result from either LER or surface impurity scattering.

\begin{figure}
\begin{center}
\leavevmode \epsfxsize=3.3in \epsffile{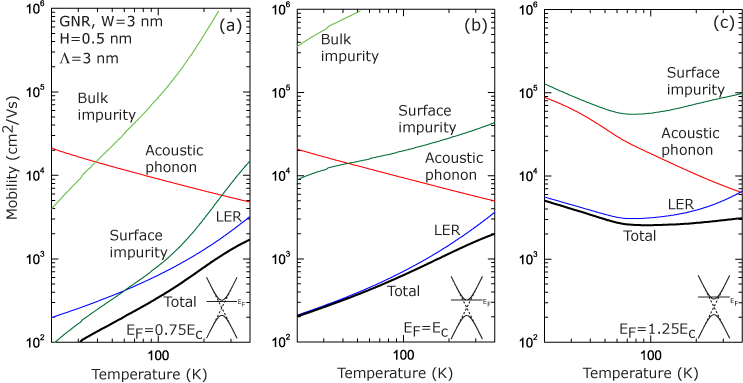}
\caption{Carrier mobility in a 3 nm GNR as a function of temperature for three distinct Fermi levels.} 
\label{mob3nm}
\end{center}
\end{figure}

If the width of the GNR is lowered to $W = 3$ nm but with a roughness the same as the 5 nm wide GNR, LER scattering becomes stronger.  Figure \ref{mob3nm} depicts the relative importance of the various scattering mechanisms at different temperature for three locations of the Fermi level.  As can be seen, LER scattering starts dominating at and around room temperature as well, overtaking acoustic phonon scattering.  If one considers the case when surface and bulk impurities are absent, the mobility will be determined by the relative importance of acoustic phonon and LER scattering.

\begin{figure}
\begin{center}
\leavevmode \epsfxsize=2.3in \epsffile{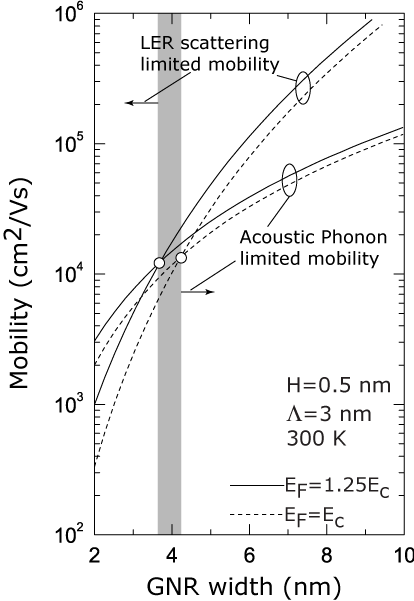}
\caption{For very narrow ribbons, LER scattering is seen to be the mobility-limiting scattering mechanism, but wider GNRs - even for widths greater than $\sim$5 nm are relatively insensitive to LER scattering.  For such ribbons, the mobility at room temperature is limited by acoustic phonon scattering.} 
\label{mobIntrinsic}
\end{center}
\end{figure}

Finally, if the surface impurity density is lowered to a level $<< 10^{10}$/cm$^{2}$, the LER and acoustic phonon scattering mechanisms are of the most interest.  Under such a condition, the mobility is plotted as a function of the GNR width for a constant edge roughness in Figure \ref{mobIntrinsic}.  The crossover of the LER and acoustic phonon scattering rates occurs at $W \sim 4$ nm for the chosen edge roughness parameters.  For narrow ribbons, LER scattering dominates, but as the width of the GNR is increased, the LER scattering rate decreases as $W^{4}$, making the GNRs of widths 5 nm and above relatively insensitive to edge roughness scattering.  As $W \rightarrow \infty$, the acoustic phonon scattering rate limit on the mobility approaches that of 2D graphene; indeed, mobilities as high as 120,000 cm$^{2}$/V.s have been recently observed in suspended 2D graphene sheets when the surface impurities were removed by current-induced annealing \cite{condmat08_kim_susp}.

\section{Conclusions \& Acknowledgements}

In conclusion, it is shown from analytical modeling that low-field carrier transport in GNRs is affected by intrinsic scattering by acoustic phonons, in addition to line edge roughness and impurity scattering.  Impurities sticking to the GNR surface or at the interface of GNRs and underlying substrates are much more deleterious than those embedded in the underlying substrate.  For very thin GNRs, the mobility is degraded by LER scattering, which reduces the mobility as the fourth power of the GNR width.  For many technologically relevant GNR widths, LER scattering is weaker than the intrinsic acoustic phonon scattering.  

Financial support from NSF awards DMR-06545698 (CAREER award),  ECCS-0802125, \& from the Nanoelectronics Research Initiative (NRI) through the Midwest Institute for Nanoelectronics Discovery (MIND) are gratefully acknowledged. 



\begin{references}
\bibitem{nofringing} Fringing fields are neglected here due to the 1D geometry of the GNR.  This approximation does not have a major impact on the dominant scattering rate determination, since bulk impurity scattering is found to be weaker than the surface impurity.
\end{references}
\end{document}